\documentclass[twocolumn,twoside,slac_two]{revtex4}
\usepackage{graphicx}
\usepackage{fancyhdr}
\begin{document}

\title{An investigation of alternative configurations of the read controllers of the Fermi LAT tracker}

%

\author{Leon Rochester for the Fermi-LAT Collaboration}
\affiliation{SLAC National Accelerator Laboratory, Stanford University, Stanford, CA 94305}

\begin{abstract}
The Fermi Large Area Telescope (LAT)~\cite{LAT_ref} consists of 16 towers, each incorporating a tracker made up of a stack of 18 pairs of orthogonal silicon strip detectors (SSDs), interspersed with tungsten converter foils. The strip numbers of the struck strips in each SSD plane are collected by two read controllers (RCs), one at each end, and nine RCs are connected by one of eight cables to a cable controller (CC).

The tracker readout electronics limit the number of strips that can be read out.  Although each RC can store up to 64 hits, a CC can store maximum of only 128 hits. To insure that the photon shower development and backsplash in the lower layers of the tracker don't compromise the readout of the upper layers, we artificially limit the number of strips read out into each RC to 14, so that no CC can ever can see more than 126 hit strips.

In this contribution, we explore other configurations that will allow for a more complete readout of large events, and investigate some of the consequences of using these configurations.

\end{abstract}

\maketitle

\thispagestyle{fancy}


\section{Introduction}

The Large Area Telescope (LAT)~\cite{LAT_ref} is the main instrument on the Fermi Gamma-ray Space Telescope, which was launched in June 2008 and has been surveying the gamma-ray sky since then. The data collected have led to numerous discoveries in this previously relatively unexplored energy regime. 

The LAT consists of 16 towers in a 4x4 array, each incorporating a tracker and a cesium-iodide-based calorimeter, all surrounded by an tiled scintillator anti-coincidence detector.

\section{Hit truncation in the tracker data buffers}

Each tracker module contains 36 planes of silicon-strip detectors, paired orthogonally in 18 layers interspersed with tungsten foils.
The hit strip numbers and associated information in each of the tracker planes are read out into buffers in two read controllers (RCs), one at each end of the plane, and nine RCs are read into a buffer in one of eight cable controller (CC), where the data are stored for assembly into the complete event. Each RC can accommodate up to 64 strips, but the CC can only accept the first 128. If there are more than 128 hits, those from the top planes are lost first, potentially leading to a serious loss of resolution on the reconstructed photon direction.

So we limit the number of hits in each RC so that the CC buffer is never completely filled. For events with many hits, this strategy tends to confine the hit loss to the lower planes, where the photon has started to shower, and where the tracker is sensitive to backsplash due to low-energy photons from the calorimeter, just below the tracker.

\begin{figure}
\includegraphics[width=80mm]{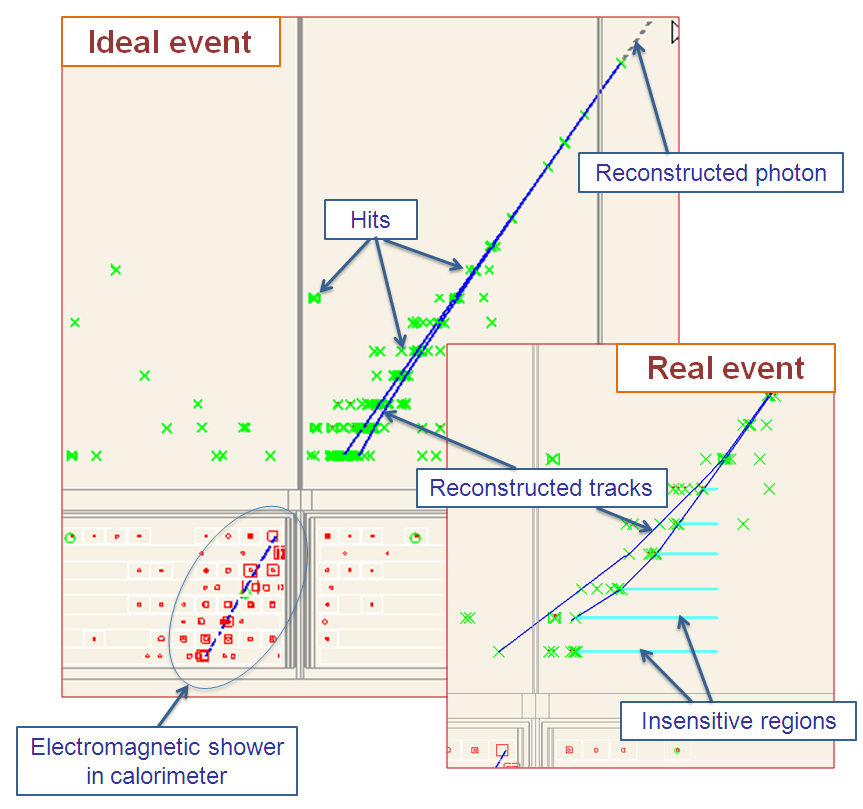}\caption{A truncated event}\label{TruncEvent}
\end{figure}

Figure~\ref{TruncEvent} shows a simulated high-energy photon showering in the tracker, with no limits on the number of hits read out. This can be done in simulation, but not in the real detector. As the shower develops, the number of hit clusters (green x's) in each layer increases. The blue lines indicate the tracks found by our new pattern-recognition algorithm~\cite{Tracy_ref}. 

The inset shows the same event, but now the hit buffers are truncated as they would be in the standard configuration of the LAT readout electronics. (See next panel.) Many of the hits in the lower layers are lost; the teal bars show the regions that become insensitive. Because of the missing hits, the lower parts of the tracks are displaced, and this contributes to a shift in the reconstructed direction of the found tracks away from the true direction, and thus in the inferred photon direction. For this event, the shift is $0.03^\circ$.

\section{A proposed configuration~\cite{config_ref} with fewer lost hits }

\begin{figure}
\includegraphics[width=80mm]{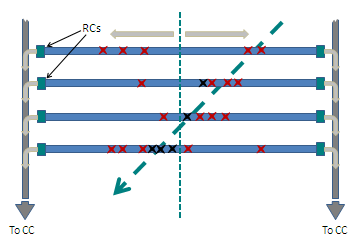}\caption{Standard readout configuration}\label{Standard}
\end{figure}

Figure~\ref{Standard} is a schematic representation of the standard configuration of the tracker readout, for one of the two orthogonal directions.  Note the two read controllers on each plane, each of which reads out the hits in half of the plane. All the planes are configured in the same way. 
In the actual readout, each read controller can accept 14 hits; here, for illustration, we set the buffer limit to three. In this event, the dark green arrow represents a photon showering in the tracker. The gray arrows indicate the direction of the information flow. The red x's are the recorded hits, and we show the lost hits in black. Note that in this event the hits towards the center of the plane are the ones affected.

\begin{figure}[floatfix]
\includegraphics[width=80mm]{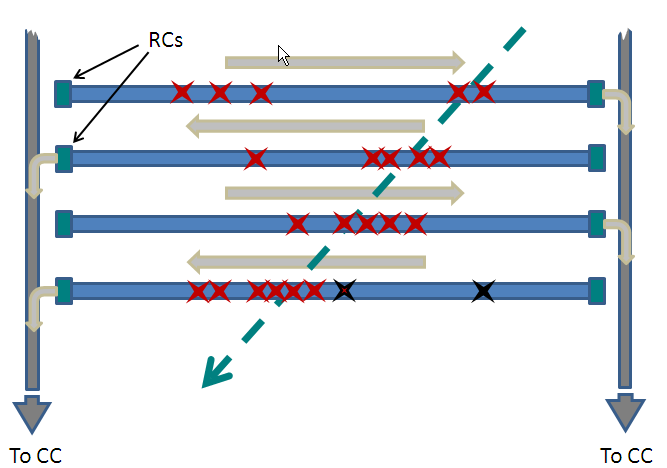}\caption{Proposed readout configuration}\label{Proposed}
\end{figure}

The proposed configuration, shown in Figure~\ref{Proposed}, takes advantage of the fact that the width of the shower in a tracker plane is typically much smaller than the half-width of the plane, so that the hits tend to fall into one half or the other. Instead of splitting the plane, we read out all the hits at one end, and double the buffer size. In the next plane, all the hits are read out at the other end. The maximum number of hits presented to each cable controller stays the same, but we now can use the currently often wasted capacity of the end farthest from the hits.

In this example, we lose only two hits in the new configuration, instead of the five in the standard. The remaining lost hits fall outside of the shower core, and will only marginally affect the tracking. This is a general feature of this configuration, since the hits are lost from the ends, rather than from the middle of the plane.
As a variation of this configuration, we can “taper” the buffer limits, so that the buffer size is smaller at the top of the tracker, where there are fewer hits, allowing us to use the extra capacity at the bottom. The example we will use below tapers from 12 hits at the top to 49 at the bottom.
Readout configurations are defined in the onboard software, and can be uploaded to the orbiting instrument.

\section{Improvement in angular resolution }

\begin{figure}
\includegraphics[width=60mm]{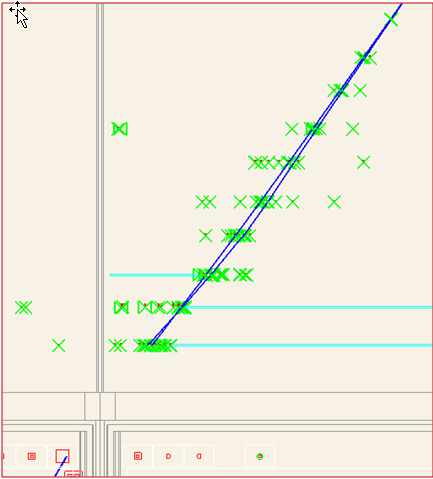}\caption{Original event with tapered readout}\label{Tapered}
\end{figure}

In Figure~\ref{Tapered}, we show the original event, read out with a tapered configuration, as described above.  Note that even though there are still lost hits, the overall situation is much closer to the ideal one.

\begin{figure}
\includegraphics[width=80mm]{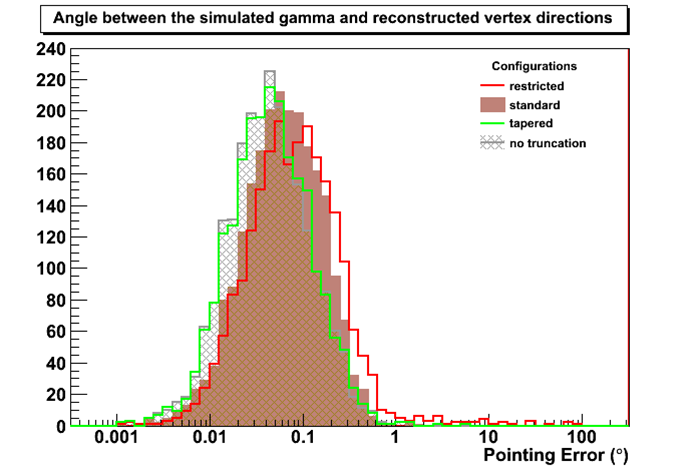}\caption{PSF for the various configurations}\label{Plot}
\end{figure}

In Figure~\ref{Plot}, we compare the angular deviations of the reconstructed tracks from the true direction (``PSF''s), for the configurations discussed above, for an event sample consisting of photons with a $1/E$ energy spectrum (uniform in $log(E)$), and incident on the LAT at $45^\circ$ to the vertical, and with a minimal cut on the overall quality of the reconstruction. We select the subsample with energies between 10 and ~300 GeV that convert in the upper (“thin radiator”) section of tracker. The energy, angle and conversion point are all chosen to maximize the truncation effect. We include one extra configuration, called ``restricted." This is similar to the standard configuration, except that the maximum number of hits in an RC is restricted to eight. This configuration allows us to compare the effects of truncation in the simulation to those in the real data, since our existing data can be truncated {\it a posteriori\/} to simulate the restricted configuration.

To make the plot, we choose all events for which the PSF for the standard configuration is below two degrees. Since there are only three events between one and two degrees, the exact location of this cut is not crucial.

\section{Quantitative results and discussion  }

To make a quantitative comparison, we take the means of the distributions in the previous panel, cutting the distributions off at five degrees. The qualitative conclusion doesn't depend on this cut. The results of this simple study, shown in Table~\ref{Results}, indicate that there is definitely something to be gained by adopting one of the proposed configurations. (Recent studies suggest that similar gains can be achieved using configurations which allow the CC buffers to overflow, but which provide a fuller reconstruction of events which originate in the lower part of the tracker.)


\begin{table}
\begin{center}
\caption{Quantitative results}
\begin{tabular}{|c|c|}
\hline \textbf{Configuration} & \textbf{Average Angular}  \\
  & \textbf{Deviation} \\
\hline Ideal & $0.071^\circ$ \\
 (no truncations)  & \\
\hline tapered & $0.074^\circ$ \\
\hline alternate 28 & $0.077^\circ$ \\
\hline standard & $0.092^\circ$ \\
\hline restricted & $0.144^\circ$ \\
\hline
\end{tabular}
\label{Results}
\end{center}
\end{table}

Before making a serious proposal for a change of configuration we need to do a more realistic study, using photons at all angles, and background events. We should also compare the simulation to real data, first by using the 
restricted configuration with existing data, and eventually by running tests on-orbit with the proposed configuration(s).
At the same time, we need to consider other effects of such a change. Some possibilities are:
degradation of the hardware trigger, principally through possible changes in timing;
decreased ability to identify out-of-time tracks (ghosts), because of loss of granularity of the detailed trigger information; and
increased time to reconstruct the events off-line, which can be significant for our combinatoric algorithm.

Finally, now that we're aware of this issue, we can tune our reconstruction algorithms to mitigate the effect of the missing hits.

\bigskip 
\begin{acknowledgments}
The Fermi LAT Collaboration acknowledges sup-
port from a number of agencies and institutes for
both development and the operation of the LAT
as well as scientific data analysis. These include
NASA and DOE in the United States, CEA/Irfu
and IN2P3/CNRS in France, ASI and INFN in
Italy, MEXT, KEK, and JAXA in Japan, and the
K. A. Wallenberg Foundation, the Swedish Research
Council and the National Space Board in Sweden.
Additional support from INAF in Italy and CNES
in France for science analysis during the operations
phase is also gratefully acknowledged.

Work supported by Department of Energy contract DE-AC03-76SF00515.
\end{acknowledgments}



\end{document}